%% file: mcdonald.tex
\documentclass[11pt,a4paper]{article}
\usepackage[left=15mm, top=10mm, right=15mm, bottom=20mm, nohead=true, nofoot=true]{geometry}

\usepackage[utf8]{inputenc}
\usepackage{authblk}
\usepackage{indentfirst}
\usepackage{misccorr}
\usepackage{graphicx}
\usepackage{amsmath}
\usepackage{amssymb}
\usepackage{multicol}
\usepackage{bm}
\usepackage{longtable}
\usepackage{pdflscape}
\usepackage{epstopdf}
\usepackage{multirow}
\usepackage{adjustbox}
\usepackage{amsfonts}
\usepackage{ifthen}
\usepackage{fancyhdr}
\usepackage{setspace}
\usepackage{xcolor}
\usepackage[round,authoryear,comma,sort&compress,numbers]{natbib}
\usepackage[toc,title,page]{appendix}
\usepackage[hidelinks]{hyperref}
\usepackage{url}

\hypersetup{
    colorlinks=true,
    linkcolor=green,
    citecolor=blue,
    filecolor=magenta,
    urlcolor=cyan,
    pdftitle={Sharelatex Example},
    pdfpagemode=FullScreen,
}

\fancypagestyle{plain}{\chead{\texttt{\textcolor{brown}{Communications of BAO, Vol. ?, Issue ?, 20??, pp. ???-???}}}}
\title{\textbf{The \emph{Gaia} All-Sky Stellar Parameters Service (GASPS)}}
\author[1]{I.~McDonald \thanks{iain.mcdonald-2@manchester.ac.uk, Corresponding author}}
\author[1]{A.A.~Zijlstra} %\thanks{B.B@university.edu}}
\author[2]{N.J.~Cox} %\thanks{C.C@university.edu}}
\author[2]{J.~Bernard-Salas} %D.D. Author\thanks{D.D@university.edu}}
\affil[1]{\scriptsize Jodrell Bank Centre for Astrophysics, University of Manchester, Ozford Road, Manchester, M13 9PL, UK}
\affil[2]{\scriptsize ACRI-ST, Av.~Nicolas Copernic, 06130 Grasse, France}


\begin{document}
\pagestyle{empty}
\newpage
\pagestyle{fancy}
\label{firstpage}
\date{}
\maketitle

\begin{abstract}
Temperature and luminosity are the two key diagnostics of a star, yet these cannot come directly from survey data, but must be imputed by comparing those data to models. SED fitting offers a high-precison method to obtain both parameters for stars where both their distance and extinction are well known. The recent publication of many all-sky or large-area surveys coincides the publication of parallaxes and 3D extinction cubes from the \emph{Gaia} satellite, making it possible to perform SED fitting of truly large ($>10^8$) numbers of Galactic stars for the first time. The analysis of this data requires a high level of automation. Here, we describe the ongoing Gaia All-Sky Stellar Parameters Service (GASPS): the fitting of 240 million SEDs from \emph{Gaia} DR3 and the extraction of temperatures and luminosities for the corresponding stars using the {\tt PySSED} code. We demonstrate the quality of the initial results, and the promise that these data show, from wavelength-specific information such as the ultraviolet and infrared excess of each star, to stellar classification, to expansion of the project beyond our own Galaxy, and mineralogical mapping of the Milky Way's interstellar medium.
\end{abstract}

\section{Introduction}
The parameters of individual stars are of vital importance in a variety of astronomical disciplines, from the study of those individual stars to the stellar populations and galaxies that they are part of. The most valuable parameters are temperature and luminosity, with which stars can be placed on the Hertzsprung--Russell diagram \citep{Hertzsprung1911} and, by comparing this diagram to evolutionary models \citep[e.g.,][]{Bressan2012}, properties such as age, mass and evolutionary stage can be derived.

There are two common methods for determining stellar temperature, using spectra or photometry \citep{Russell1914}. Traditionally, spectral types in the \citet{Morgan1943} system were assigned by comparing observations to standard spectra. These were often then plotted directly on the H--R diagram, though look-up tables could also be used to convert to effective temperature. Alternatively, the difference in magnitude in two filtered observations (e.g., $B-V$) could be used to generate a colour--magnitude diagram, and stellar temperatures estimated from direct mathematical derivation from the Planck blackbody function.

These ways of generating stellar temperature are limited. In particular, both spectral type and colour depend not only on temperature, but also the surface gravity and composition of the star \citep[e.g.,][]{Kurucz2005}. Modern methods of temperature determination therefore rely on updated versions of these approaches. Detailed fitting of individual spectral lines can provide effective temperature ($T$), surface gravity ($g$), detailed composition and velocity information \citep[e.g.,][]{Cordero2014}. By assuming a mass, $M$, these can be converted into a luminosity via the relation $L \propto T^4 M/g$. The accuracy of this approach is generally limited by two factors: the accuracy of the assumed mass and the accuracy of the surface-gravity determination. Typical errors for all-sky or very-large-scale surveys might be $\Delta \log(g) \approx 0.1$ dex, or $\Delta g \approx 26$\% \citep{Avdeeva2024}, sometimes with comparable errors in mass. While these uncertainties depend crucially on the quality and coverage of the spectra, they generally impart a similarly large uncertainty in luminosity, and on derived parameters such as stellar age \citep[e.g.,][]{Kordopatis2023}.

The advanced photometric alternative is to replace colour--temperature relations with spectral-energy-distribution (SED) fitting. This approach takes synthetic photometry (derived from the same set of stellar atmosphere models as used in spectral methods) and fits it to multi-wavelength photometric observations of stars. SED fitting provides a temperature by fitting the position of the SED peak, and an angular source size by scaling the flux in the Rayleigh--Jeans tail \citep{Blackwell1977}. If a distance is known, this angular size can be converted to a  physical size, and a luminosity found via the Stefan--Boltzmann law \citep{Boltzmann1884}. There are relative advantages and disadvantages to this method. SED fitting has the advantage that it can normally be done cheaply from existing astronomical surveys, without the requirement to obtain new spectra. However, it has the disadvantages of needing a distance and requiring a photometric correction for the effects of interstellar dust. This makes SED fitting advantageous to fainter sources where taking spectra is impractical, but where those sources still have a known distance and reddening (such as faint stars with parallaxes, or stars in external clusters and galaxies). It can also provide a measurement that is largely independent of spectroscopic determinations, allowing a check on their accuracy. SED fitting is also a more robust determinant of stellar parameters in cases where stellar surfaces are not in local thermodynamic equilibrium, as is the case for some variable stars, or where spectra are too complex to model, such as for some cool stars \citep{Lind2024}.

SED fitting can also provide deviations from a stellar atmosphere model as a function of wavelength. While these deviations can be in individual bands, they more commonly manifest as a global ultraviolet or infrared excess for each star. Ultraviolet excesses generally trace a hot companion (normally either an accreting star or a young white dwarf; \citet[e.g.,][]{Montarges2025}), while infrared excess generally traces cooler material in the circumstellar environment caused by mass loss or accretion \citep[e.g.,][]{Hoefner2018}. Independent of stellar-parameter estimation, SED collation and comparison are therefore useful in identifying classes of mass-losing/accreting stars, allowing exploration of the cosmic cycle of matter in a statistical sense \citep[e.g.,][]{Boyer2011}, especially when used alongside spectroscopic tracers \citep[e.g.,][]{Jones2017}.

Previous SED-fitting analyses involving members of our group have concentrated on Galactic globular clusters \citep{Boyer2009,McDonald2009,McDonald2010,McDonald2011}, where comparison with spectroscopic estimates indicate that absolute accuracies in temperature can be as low as $\Delta T \pm 50\,\mathrm{K}$, with luminosity errors limited to a few percent by the distance determination to the individual cluster \citep{McDonald2011b}. When extended to other galaxies in the Local Group, larger errors occur due to poorer-quality photometry and a wider range of possible stellar masses and metallicities \citep[e.g.,][]{Boyer2015,Boyer2025}.

Recent years have seen a proliferation of all-sky surveys that can provide the raw photometry for SED fitting (see, e.g., A.~Mickaelian, this volume). However, progress in understanding the stellar component of our own Galaxy has been limited by the need for distances and reddening corrections. Early efforts in SED-based parameter estimation were limited to the $\sim$100\,000 \emph{Hipparcos} stars \citep{McDonald2012} and later the $\sim$1.5 million \emph{Tycho}--\emph{Gaia} cross-correlated catalogues in \emph{Gaia} Data Release 1 \citep{McDonald2017}. Since then, \emph{Gaia} has provided a revolution in distance estimation for Galactic stars, with a substantial minority of the $1.8 \times 10^9$ objects in its Data Release 3 \citep[DR3;][]{GaiaDR3} having statistically significant parallax measurements. Such measurements have also allowed the construction of 3D extinction cubes \citep{Lallement2022}, providing the two data sources needed to accurately derive SED-based properties of Galactic stars. It is therefore timely to embark on a determination of SED-fitted stellar parameters for all \emph{Gaia} DR3 sources with determined parallaxes. On this basis, the \emph{Gaia} All-Sky Parameters Service (GASPS) was conceived.

\section{GASPS}

GASPS aims to collate the SEDs of all objects in \emph{Gaia} DR3 by cross-referencing the \emph{Gaia} source list with other catalogues, and fitting those sources with secure parallax-derived distances. To perform this task effectively, we use the Python-based software package {\tt PySSED} version 1.2 \citep[][and in prep.]{McDonald2024,McDonald2025}, which is designed for fast, automated collection of SEDs from published catalogue data accessed at the Centre de Donn\'ees astronomiques de Strasbourg (CDS\footnote{\url{https://cds.unistra.fr/}}) through the {\tt astroquery} tool \citep{Ginsburg2019}. {\tt PySSED} solves several crucial problems inherent in catalogue cross-matching, namely those of one-to-many and many-to-one associations, variable beam sizes, proper-motion correction, bad-data identification (e.g., due to quality flags, saturation, etc.), and outlier rejection. By default, it uses the \citet{Lallement2022} 6\,kpc $\times$ 6\,kpc extinction cube to deredden sources.

The {\tt BT-Settl} models \citep{Allard2014} were chosen as the model dataset, largely because it offers an essentially complete grid over a wide range of temperatures, gravities and metallicities. The choice of models has a relatively small effect on the parameters produced, as SED fitting is generally only sensitive to changes in flux over large-scale spectral regions. Factors such as incompleteness of line lists therefore only become a significant problem when their absence creates significant changes to the spectra, at which point secondary effects (e.g., stellar activity, out-of-equilibrium physics) typically become more important. Each model spectrum in the grid is convolved with the filter functions of the 26 queried catalogues (see below), and reddening estimates ($A_\lambda / A_V$) generated at $A_V = 0.1$, 3.1, 10 and 31 magnitudes. These are then saved as pre-computed look-up tables. The set of tables at different $A_V$ are required due to the finite width of each filter bandpass, thus the variable amount of reddening between their blue and red extrema.

To process this large dataset, the sky was divided up into strips in Galactic latitude of $1^\circ$, and then in Galactic longitude at approximately $1^\circ$ spacing. Due to the density in the Galactic plane, fields at $|b| < 15^\circ$ were sub-divided into $0.5 \times 0.5^\circ$ fields, while fields in the range $-9^\circ \leq b \leq +6^\circ$ were sub-divided into $0.2 \times 0.2^\circ$ fields. In each field, the \emph{Gaia} catalogue was cross-matched against 25 other catalogues, including the combined ultraviolet data from the \emph{GALEX} satellite \citep{GALEX}; optical catalogues from \emph{Hipparcos}/\emph{Tycho} \citep{Perryman97}, PanSTARRS DR1 \citep{PanSTARRS-DR1}, APASS DR9 \citep{HLTW15}, SDSS DR16 \citep{SDSS16}, Skymapper DR4 \citep{Onken2024}, CMC15 \citep{Muinos2014}, DES DR2 \citep{DESDR2}, TASS \citep{DRSC06}, and the surveys of the VST \citep{Drew14,Shanks15}; near-infrared surveys from 2MASS \citep{SCS+06}, UKIDSS \citep{LWA+07,Lucas2008} and the VISTA telescope \citep{VHS,Smith2025}; and various mid-infrared catalogues from the \emph{Akari} \citep{Akari}, \emph{Spitzer} \citep{Churchwell2009,SSTSL2} and \emph{WISE} satellites \citep{AllWISE,catWISE,unWISE}. A bespoke search radius is employed for each catalogue, set conservatively to minimise false positives.

Magnitude limits for each catalogue are set to avoid saturation, loss of linearity and excess photometric noise, while additional quality flags are employed to remove known problems with specific catalogues, or to choose certain catalogues (e.g., different \emph{WISE} reductions) over others given certain criteria (such as limiting magnitude). Many of these selections were inherited from the cuts employed by \citet{McDonald2017}. When multiple observations in the same photometric filter are available, these are merged if they agree within errors, or the more precise measurement chosen if they do not. A noise floor is also added to each filter in each catalogue: systematic uncertainties (e.g., from zero-point errors, source blending or intrinsic variability) normally dominate over photon-based random uncertainties in large catalogues (see \citet{McDonald2024} for discussion). Typically this noise floor is 10\% for optical, 20\% for near-infrared and 25--33\% for mid-infrared observations.

Having collected SEDs for each object, {\tt PySSED} attempts to identify a distance based on (in order of preference) the geometric distances of \citet{BailerJones2021}, direct inversion of \emph{Gaia} DR3 parallaxes, and direct inversion of \emph{Hipparcos} parallaxes. From this distance, an extinction is drawn from Lallement et al.'s cube and used to deredden the photometry. Next, a stellar composition and surface gravity are needed to select the correct family of stellar models. Abundances for iron and $\alpha$-elements are taken either from \emph{Gaia} $B_P$ and $R_P$ spectra from \emph{Gaia} DR3 or, if these are unavailable, from the following estimate:
\begin{equation}
{\rm [Fe/H]} = \tan^{-1} Z , \qquad {\rm [\alpha/Fe]} = \frac{\tan^{-1} Z}{5} ,
\end{equation}
An initial estimate for the surface gravity is generated by fitting a blackbody function, which allows determination of a stellar radius and a consequent luminosity, and generates a surface gravity after assuming a stellar mass. This mass is either based on the zero-age main sequence \citep{PARSEC} for main-sequence stars, 1\,M$_\odot$ for giant stars ($T < 5500\,{\mathrm K}$) below the red giant branch (RGB) tip ($L \approx 2500\,{\mathrm L}_\odot$), or an average of the core mass from \citet{Bloecker1993} and the initial mass from \citet{Casewell2009} for mass-losing asymptotic giant branch (AGB) stars above the RGB tip. The blackbody function also provides a starting estimate of the effective temperature. While this method therefore inherits the problems from spectrally derived parameter estimates (namely requiring an assumed stellar mass), the final parameters are largely insensitive to the log($g$) and composition selected, typically generating $<$100\,K error in temperature, even for relatively large departures from the true composition and gravity \citep{McDonald2017,McDonald2024}.

Once an initial set of parameters is chosen, an appropriate stellar atmosphere model is interpolated from the pre-computed tables and compared against the data. A reduced $\chi^2$ ($\chi^2_{\rm r}$) is then computed between the model and data: this is done in logarithmic flux to equalise effects from fluxes that are too faint and too bright. A Nelder--Mead function was then used to minimise this $\chi^2_{\rm r}$ to obtain a best-fit temperature. Following this, a new surface gravity is obtained and the data refitted. The data are then tested for outliers, with the SED being refit after each outlier is removed. Data are rejected if their removal improves the $\chi^2_{\rm r}$ by an average factor of 1.5 for each removed datapoint. Up to five outliers can be treated together (without any being necessarily rejected) before the SED fit is declared stable. In practice, this limit is fairly conservative at removing outliers. If a fit converges outside the range of the stellar models, then a blackbody fit is used instead: this naturally provides a poorer fit and a less accurate set of parameters than fitting a stellar model atmosphere.

Following this process, the final results are calculated and saved to file. {\tt PySSED}'s fitting operates at a speed of around one second per star. The entirety of \emph{Gaia} DR3 required 8.6 CPU years of processing, and resulted in 2.8 TB of data (this reduces to 520 GB if only fitted sources are retained). At the time of writing, data are being re-reduced as routines have now been better optimised to remove poor-quality data.

\section{Results}

\begin{figure}[ht]
\centering
\includegraphics[width=0.97\textwidth]{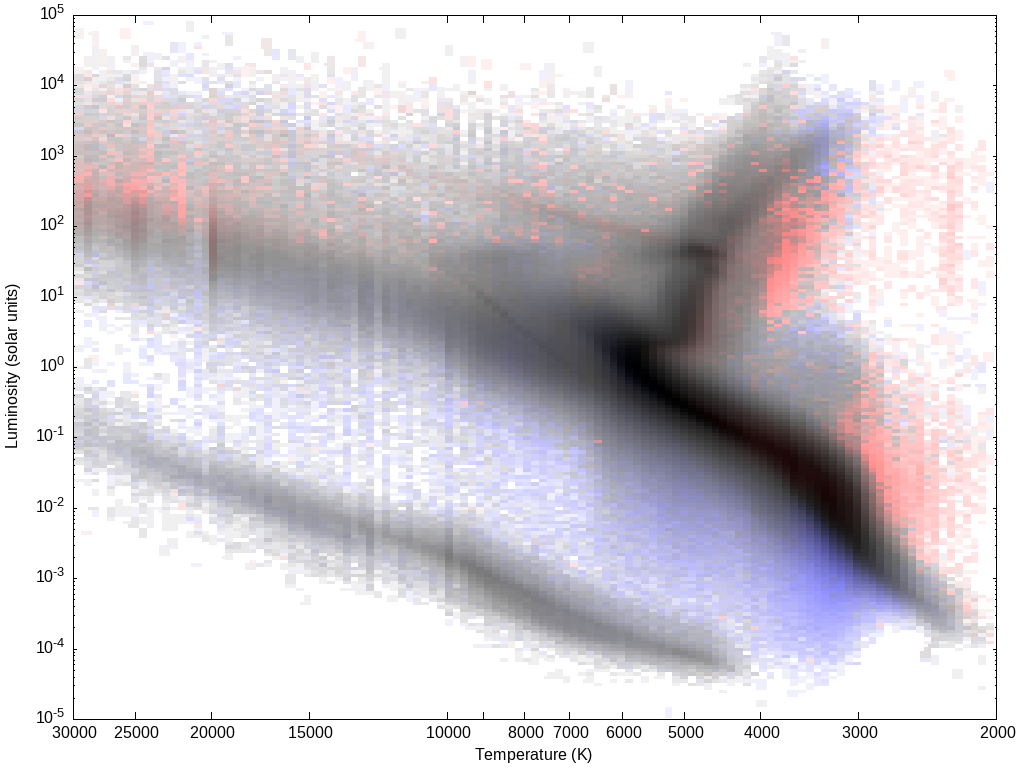}
\caption{A binned Hertzsprung--Russell diagram of sources in GASPS. The plot contains 33 million sources covering the (mostly low extinction) regions near the Galactic Poles. Point shading is proprtional to the logarithm of the number of sources in each bin to highlight areas with fewer stars (i.e., regions with problems and/or rarer stellar types). Bin colour is dictated by the average ratio of observed to modelled flux in the \emph{Gaia} $G$ band (blue is 0.8, red is 1.25). Similar plots are available for other passbands, and serve as a useful analytical tool of the accuracy of calibration in each survey and of the GASPS reduction overall.
\label{fig:HRFig}}
\end{figure}

The resulting catalogue contains the stellar parameters of $\sim$240 million objects, or around 13.3\% of \emph{Gaia} DR3 sources. The primary limitation to completeness is the measured distances of objects (limited to those with $\Delta d / d < 0.25$), with the existence of enough photometry to form a meaningful SED ($\geq$4 data points) being a secondary concern. The primary limitation to accuracy is the extinction correction: as we rely on a coarse (50\,pc) grid with its own uncertainties, the accuracy of the extinction correction can be limited by the grid resolution and accuracy. This has a particularly strong effect for regions where substantial extinction is mapped close to the Sun.

Figure \ref{fig:HRFig} shows the H--R diagram for sources near the Galactic poles, reflecting those sources re-reduced at the time of writing. This region is mostly at very low extinction, thus tests the quality of the fitting and analysis of {\tt PySSED}. The expected features of the diagram can be seen, including the main sequence, which extends from O-type stars at the left to L-type stars at the bottom-right. A prominent main-sequence turn-off can also be seen (older populations of stars are mostly seen at these latitudes), as well as the giant branch and the white-dwarf cooling track. The giant branch is noticeably spread into metal-poor (left-hand side) and metal-rich (right-hand side) components: the horizontal position of a star in the giant branch correlates well with both the \emph{Gaia} metallicity and distance above the Galactic Plane, thus demonstrating that this is a real effect. A well-defined RGB tip can also be seen, with a few AGB stars above it.

Concentrations of stars are also visible on the giant branch, including the red clump of core-helium-burning stars at around 50\,L$_\odot$. A faint horizontal branch can be seen extending off this to higher temperatures, and is particularly visible between 6000 and 10\,000\,K. Separately, the RGB bump (caused by a drop in luminosity when the hydrogen-burning shell crosses a previous mixing boundary) can be seen at a luminosity of around 30\,L$_\odot$, and an AGB bump (caused by the transition to shell helium burning) can faintly be seen around 120 L$_\odot$.

\begin{figure}[ht]
\centering
\includegraphics[width=0.47\textwidth]{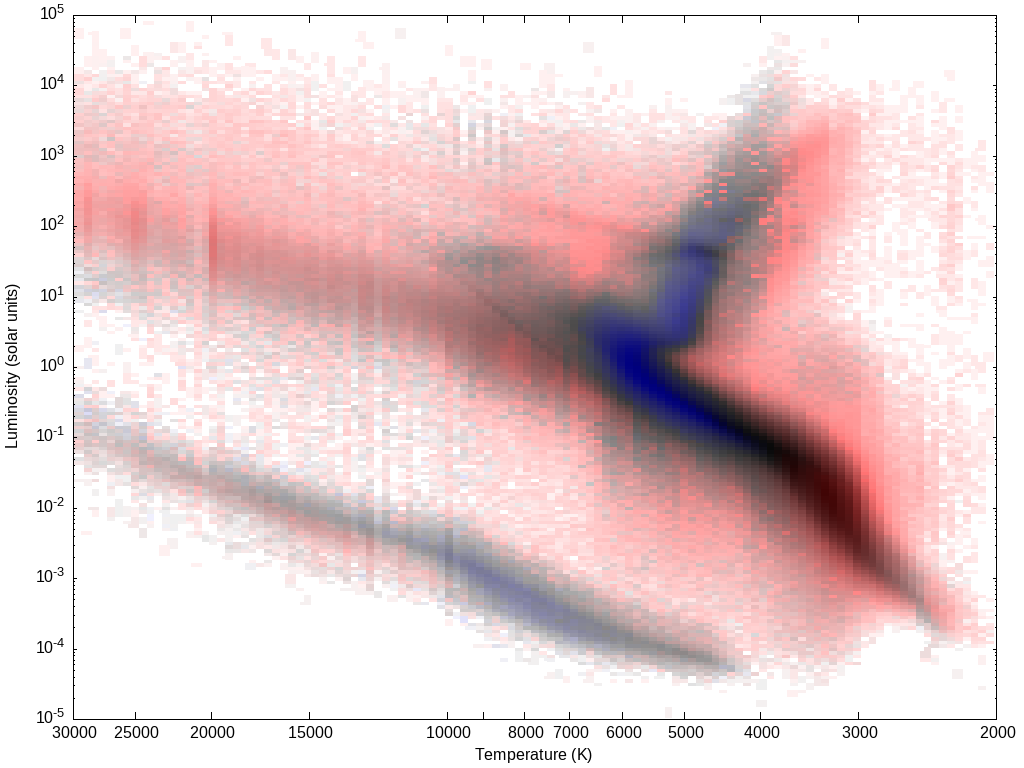}
\includegraphics[width=0.47\textwidth]{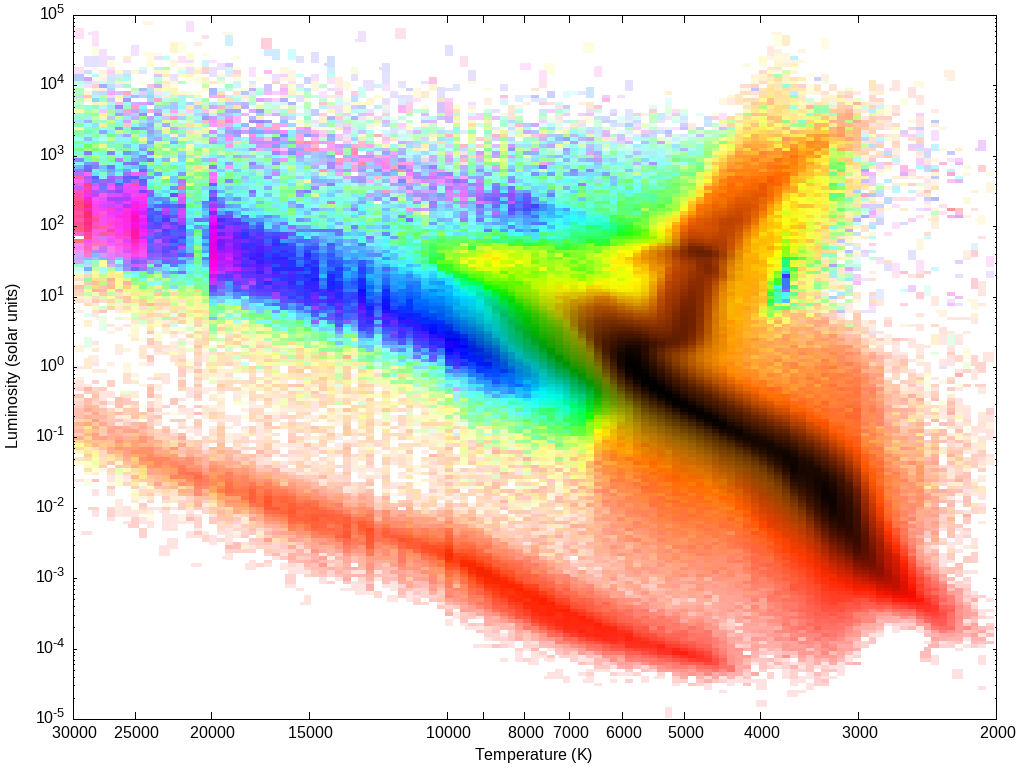}
\caption{A binned Hertzsprung--Russell diagram of sources in GASPS. As Figure \ref{fig:HRFig}, but colour-coded to show average goodness-of-fit (left, coloured from blue [0.01] through grey [0.03] to red [0.1]) and extinction (right, coloured by hue from red [$E(B-V)=0$\,mag] through green [0.33] and blue [0.67] back to red [1\,mag]).
\label{fig:HRFig2}}
\end{figure}

Some spread can be seen away from these main features. This spread appears prominently here due to the logarithmic shading and generally correlates with an excess or deficit of flux in \emph{Gaia} $G$ compared to the fitted model atmosphere. Figure \ref{fig:HRFig2} shows some of the diagnostic plots used to identify real and artificially induced features in these plots. The goodness-of-fit statistic in the left panel shows the median of the ratio of observed to modelled flux: blue and grey sources along the main sequence and white-dwarf cooling track are generally well-fit; redder sources scattering from these regions are generally poorly fit.

These poorly fit sources demonstrate a diagonal ``smearing'' across Figure \ref{fig:HRFig2}, which can be seen streaming from the top-left of both the upper main sequence and the top-left of the red clump: the right-hand panel of Figure \ref{fig:HRFig2} shows that these stars are modelled to have high extinctions. In these cases, that reddening correction appears to be over-estimated by the model cubes, likely because the stars have uncertain distances and in reality lie just in front of the extinction structures being imposed (a smaller counterpart can be seen on the opposite side, which becomes more prominent in high-extinction regions in the Galactic Plane). Brighter, bluer stars are worst affected as they remain optically visible (and with measured parallaxes) out to larger distances and thus higher average extinctions.

For fainter stars, stars spreading away from the main sequence are modelled to be at low extinction. Spread above the main sequence instead appears to be mostly due to remaining bad or noisy data in SEDs (e.g., due to remaining stellar blending or noise in faint sources). This will be true of stars below the main sequence too, although these can also be white-dwarf--main-sequence binary stars, where each component dominates a different part of the SED. These stars typically present with excess flux in the ultra-violet.

Stars near the tip of the RGB and beyond into the AGB also receive poor fits. These stars show a poor goodness-of-fit because of their intrinsic variability and dust production. These stars show excess flux in the infrared due to their warm circumstellar dust. Together, these sub-samples of ultra-violet- and infrared-excess stars can be used to examine candidate young stars, evolved stars and binary stars, for example to investigate mass-losing AGB stars, young stellar objects, Herbig Ae/Be stars or white-dwarf--main-sequence binaries.

Two further artificial effects can be seen. Vertical banding in the diagram is caused by incomplete photometry near the SED peak or noisy data, where a $\chi^2$ minimum is found exactly on a grid point of the atmospheric models: this is most prominent in poor-quality fits caused by incorrect extinctions. A diagonal feature can also be seen around 8000\,K on the main sequence. There is related to a concentration of stars with \emph{Gaia} spectrally derived temperatures at 8000\,K, which propagates to a fraction of stars in our data, since this temperature is used as a starting point for the fit and is also a temperature in our model grid.

The full fitted photometry and final parameters for each star will be released as part of the final paper, and will be available for visualisation and query via an online service, which is being created specifically for this purpose for integration with Virtual Observatory tools.

\section{Uses and future work}

We anticipate a wide variety of uses for the GASPS service. Our previous catalogues have found uses in diverse areas. These include the exoplanet community, where data are used directly as properties of exoplanet host stars \citep[e.g.,][]{AThano2023}, to select targets for follow-up \citep[e.g.,][]{Morgan2019,Hayes2024}, or to estimate secondary properties like the habitable zone of planets \citep[e.g.,][]{Chandler2016}. Other stellar communities have also used them for constructing wider datasets \citep[e.g.,][]{Antoci2019} and work on individual stars \citep[e.g.,][]{Roederer2018,Driessen2022,Wesson2024}. However, their biggest achievement above similar datasets has been that these parameters have been linked to infrared excess, which has made them highly useful in the field of dusty objects \citep[e.g.,][]{Rebull2015}, and particularly evolved stars \citep[e.g.,][]{McDonald2018}.

The first application of the {\tt PySSED} software has therefore been to create a catalogue of evolved stars within 300 pc of the Sun, and compare them against those from the Nearby Evolved Stars Survey (NESS; \citealt{Scicluna2022}). This analysis \citep{McDonald2025} allows us to measure star-formation histories of different parts of our Galaxy \citep[to be compared with external galaxies, e.g.,][and this volume]{Abdollahi2023} and derive the bias of existing evolved-star publications towards particular kinds of objects, e.g., near-infrared-bright stars with strong mass loss. It also allows NESS to be compared to other narrower but deeper modern surveys, including ATOMIUM \citep[``ALMA tracing the origins of molecules in dust forming oxygen rich M-type stars'',][]{Gottlieb2022} and DEATHSTAR \citep[``Determining accurate mass-loss rates of thermally pulsing AGB stars'',][]{Ramstedt2020}. This test demonstrates the power of GASPS to contextualise existing field-specific samples of stars using more complete statistical samples and bring together multiple projects in the field. 

Another promising aspect of GASPS is its use in stellar classification. A wide variety of stellar classifiers exist, but they typically have one of three problems: they either are based only on spectra, thus miss classes not represented in spectral features \citep[e.g.,][]{ElKholy2025,EilatBloch2025}; and/or they are specific to the confines of one survey and cannot be applied outside that survey's spatial coverage or magnitude range \citep[e.g.,][]{Woods2011,Ghaziasgar2025}, and/or they are specific to particular sub-classes of stars \citep[e.g.,][]{AwangIskandar2020,Zhou2025,GarciaZamora2025}, and thus are limited in the stellar classes they can predict. However, the full taxonomy of stars is very broad\footnote{See, e.g., https://simbad.cds.unistra.fr/guide/otypes.htx} and requires a variety of astrophysical techniques to fully place stars into all classes.

An SED alone equally does not provide data on all stellar classes, but the nature of the cross-matched data in GASPS allows it to be quickly and effectively matched against other surveys. These include variability and spectral information from \emph{Gaia}, but also features at other wavelengths that can be used to classify various young, evolved and background objects \citep[e.g.,][]{Ruffle2015,Jones2017}. Looking to future work, the in-built cross-matching of {\tt PySSED} can allow cross-matching against arbitrary catalogues to merge data from a variety of heterogeneous sources.

A pool of data is not enough either. A robust stellar classifier needs to be accurate and effective. Machine learning offers an automatable means of enhancing stellar classification. However, multi-catalogue data is by nature a sparse dataset, with many missing values. Often it is the departure from typical values that highlights a star as being in a certain class, so data imputation cannot easily be used. We therefore require machine-learning algorithms that can cope with sparse datasets, such as {\tt XGBoost} \citep{Chen2016}. {\tt XGBoost} has demonstrated the ability to deal with stellar classification of SEDs in this manner \citep{Cody2024}, but this implementation lacks accuracy because the classes were not optimised for the information contained in the SED and no information outside the SED was considered. It can be expected that training this method on specific sub-classes of stars and/or adding data from non-photometric sources can create a much higher-accuracy general classifier for stars.

Finally, we can look to improve our data coverage and data accuracy by increasing the number of stars with distances, and the robustness of those distances and the corresponding interstellar reddening corrections. All three of these will be improved by future \emph{Gaia} Data Releases, which will bring an increase in both the number of stars with parallaxes and the number of parallaxes. However, an aim of GASPS is to facilitate comparisons between easily studied samples of stars in the Milky Way and a more diverse population in other environments \citep[e.g.,][]{McQuinn2017,Gholami2025}. With our measurements of infrared excess, this specifically includes the evolution of dust throughout Universal history \citep[e.g.,][]{Boyer2017,Ghaziasgar2025b}. Consequently, future development of the GASPS project aims to focus on an identical reduction of data for stars without direct parallax distances, including Galactic globular clusters and resolved galaxies in the Local Group and Local Volume, where both distance and Milky-Way reddening are well known. This targetted approach also allows us to add data from other satellites and surveys \citep[e.g.,][]{Saremi2020,Hunt2025,Massari2025}.

We have several further options to increase the accuracy of our reddening corrections. \emph{Gaia} DR4 will allow the construction of larger and more precise data cubes of extinction. However, if SEDs have enough wavelength coverage, it should be possible to fit optical extinction (e.g., $A_V$) and temperature simultaneously, thus removing the need for indirect extinction corrections. While initial results are promising, further work is needed to extend the locally successful results to the entire sky \citep{Csukai2025}. If data are good enough, then not only can total extinction be measured, but the slope of the extinction curve ($R_\lambda = A_\lambda / A_V$) can be fit. Since $R_\lambda$ depends on the properties of interstellar dust, this offers our first opportunity to make a mineralogical map of the Milky Way's dust.

\section{Conclusions}

The GASPS project can therefore act as a key fulcrum in the meta-analysis of astrophysical data, allowing us to properly form a census of stars in the Milky Way. As well as measurement of stellar parameters and infrared excess, we can hope to extend this catalogue to include ancillary information from other surveys and classify stars, finding shining examples and common inhabitants of all classes. Finally, we can look beyond the Milky Way as a collection of indvidiual stars, and start to treat it as a statistical sample that can be accurately placed in the context of the galaxies around us and therefore the wider Universe, with its rich history and future.

%\clearpage % To force this stuff to happen by this point in the text, otherwise these will probably end up after the references.

\section*{\small Acknowledgements}
\scriptsize{The authors acknowledge funding from OSCARS. The OSCARS project has received funding from the European Commission’s Horizon Europe Research and Innovation programme under grant agreement No. 101129751.}

\scriptsize
\bibliographystyle{ComBAO}
\nocite{*}
\bibliography{references}

\end{document}